\documentclass[11pt,twoside]{article}
\usepackage{graphicx,epsfig,natbib,epstopdf}
\usepackage{CS18}
\begin{document}
%
%
%
\title{On the Spectroscopic Properties of the Retired A Star HD 185351}
%
%
\author{Luan Ghezzi$^{1}$, Jos\'e Dias do Nascimento$^{1,2}$, John Asher Johnson$^{1}$}
\affil{$^1$Harvard-Smithsonian Center for Astrophysics, 60 Garden Street, Cambridge, Massachusetts 02138 USA; lghezzi@cfa.harvard.edu}
\affil{$^2$Universidade Federal do Rio Grande do Norte (UFRN), Dep. de F\'isica Te\'orica e Experimental (DFTE), CP 1641, 59072-970 Natal, RN, Brazil}
\begin{abstract}
%
%
Doppler-based planet surveys have shown that, besides metallicity, the planet occurrence is also correlated
with stellar mass, increasing from M to F-A spectral types. However, it has recently been argued that the
subgiants (which represent A stars after they evolve off the main sequence) may not be as massive as
suggested initially, which would significantly change the correlation found. To start investigating this claim, we
have studied the subgiant star HD 185351, which has precisely measured physical properties based on
asteroseismology and interferometry. An independent spectroscopic differential analysis based on excitation
and ionization balance of iron lines yielded the atmospheric parameters $T_{\rm eff}$ = 5035 $\pm$ 29 K, $\log$ g = 3.30 
$\pm$ 0.08 and [Fe/H] = 0.10 $\pm$ 0.04. These were used in conjunction with the PARSEC stellar evolutionary tracks
to infer a mass M = 1.77 $\pm$ 0.04 M$_{\odot}$, which agrees well with the previous estimates. Lithium abundance
was also estimated from spectral synthesis (A(Li) = 0.77 $\pm$ 0.07) and, together with $T_{\rm eff}$ and [Fe/H], 
allowed to determine a mass M = 1.64 $\pm$ 0.06 M$_{\odot}$, which is independent of the star's parallax and surface gravity. 
Our new measurements of the stellar mass support the notion that HD185351 is a Retired A Star with a mass in excess of 
1.6 M$_{\odot}$.
\end{abstract}
%
%
%
%
%
\section{Introduction}

It is well-known that giant planet occurrence is strongly dependent on stellar
metallicity for main-sequence and subgiant stars (e.g., \citealt{2005ApJ...622.1102F}). Recent
works suggest that planet occurrence is also correlated with stellar mass, with the probability 
of a star hosting a Jovian planet rising linearly over the mass range 0.2 to 2.0 M$_{\odot}$ 
(e.g., \citealt{2010PASP..122..905J}). However, it has recently been argued
that the subgiants/giants studied at the massive end off the planet-host mass distribution
(representing A stars after they evolve off the main sequence) may not be as massive
as initially suggested \citep{2011ApJ...739L..49L,2013ApJ...774L...2L,2013ApJ...772..143S}.
If the stellar masses have been previously underestimated, the nature and interpretation of the 
correlation between planet occurrence and stellar mass would need to be revised. 
An accurate assessment of the stellar
masses of subgiants and giants is thus required to fully understand this apparent
correlation. In this work, we start to address this question by
performing a spectroscopic analysis of the giant/subgiant star HD 185351.

\section{Observations and Analysis}

Three spectra were obtained for HD 185351 using the High Resolution Echelle
Spectrometer (HIRES; \citealt{1994SPIE.2198..362V}) on the Keck I 10-meter telescope 
(Mauna Kea, Hawaii). They have a nearly complete wavelength coverage from 3650 to 
7950 \AA, resolution R = 77,000 and S/N $\sim$ 250 per resolution element at $\sim$6700 \AA.

We measured the atmospheric parameters
($T_{\rm eff}$, $\log$ g, [Fe/H] and $\xi$) in LTE using an
iterative method based on the excitation
and ionization equilibria of Fe I and Fe II
lines, ARES \citep{2007A&A...469..783S}, MOOG
\citep{1973PhDT.......180S} and Kurucz ODFNEW model
atmospheres \citep{2004astro.ph..5087C}. The
results were: $T_{\rm eff}$ = 5035 $\pm$ 29 K, 
$\log$ g = 3.30 $\pm$ 0.08, [Fe/H] = 0.10 $\pm$ 0.04 and $\xi$ = 1.08
$\pm$ 0.03 km s$^{-1}$.

\begin{figure}
\plotone{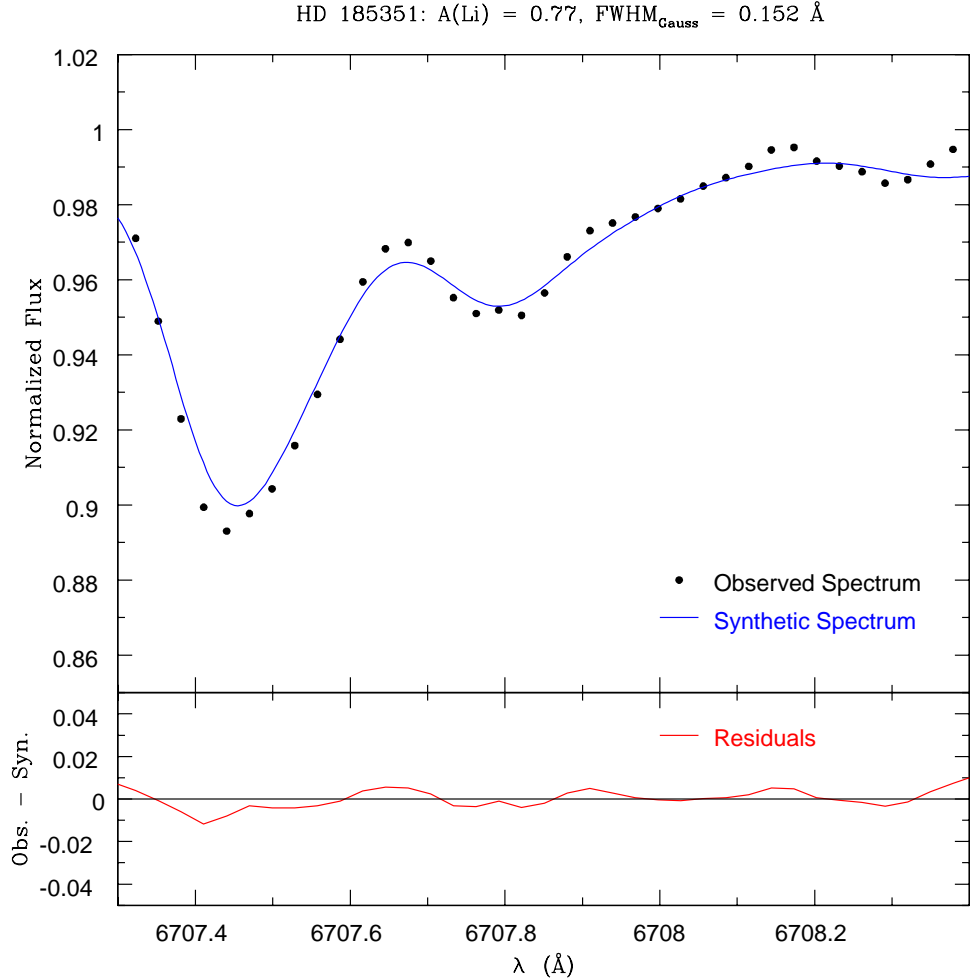}
\caption{\textit{Top panel:} Best fit between the observed (black points) and synthetic 
(blue line) spectra for HD 185351 in the region around the Li I line ($\sim$6708 \AA).
The lithium abundance and Gaussian broadening used to construct the synthetic spectrum are 
shown on the top of the figure. \textit{Bottom panel:} Residuals of the fit (red line), showing 
an overall good agreement.}
\label{synth_li}
\end{figure}

We also measured the lithium abundance in
LTE using a spectral synthesis technique, the
atmospheric parameters above and MOOG.
Our iterative procedure (see \citealt{2010ApJ...724..154G} for details) 
tested different combinations for A(Li) and a single Gaussian
broadening FHWM$_{Gauss}$. A reduced $\chi^{2}$
minimization was used to find the best fit
between the observed and synthetic spectra, which occurred for
the following parameters: A(Li) = 0.77 $\pm$ 0.07 and FHWM$_{Gauss}$ = 0.152 \AA\,
(see Figure \ref{synth_li}).

\section{Preliminary Results and Discussion}

\begin{figure}
\plotone{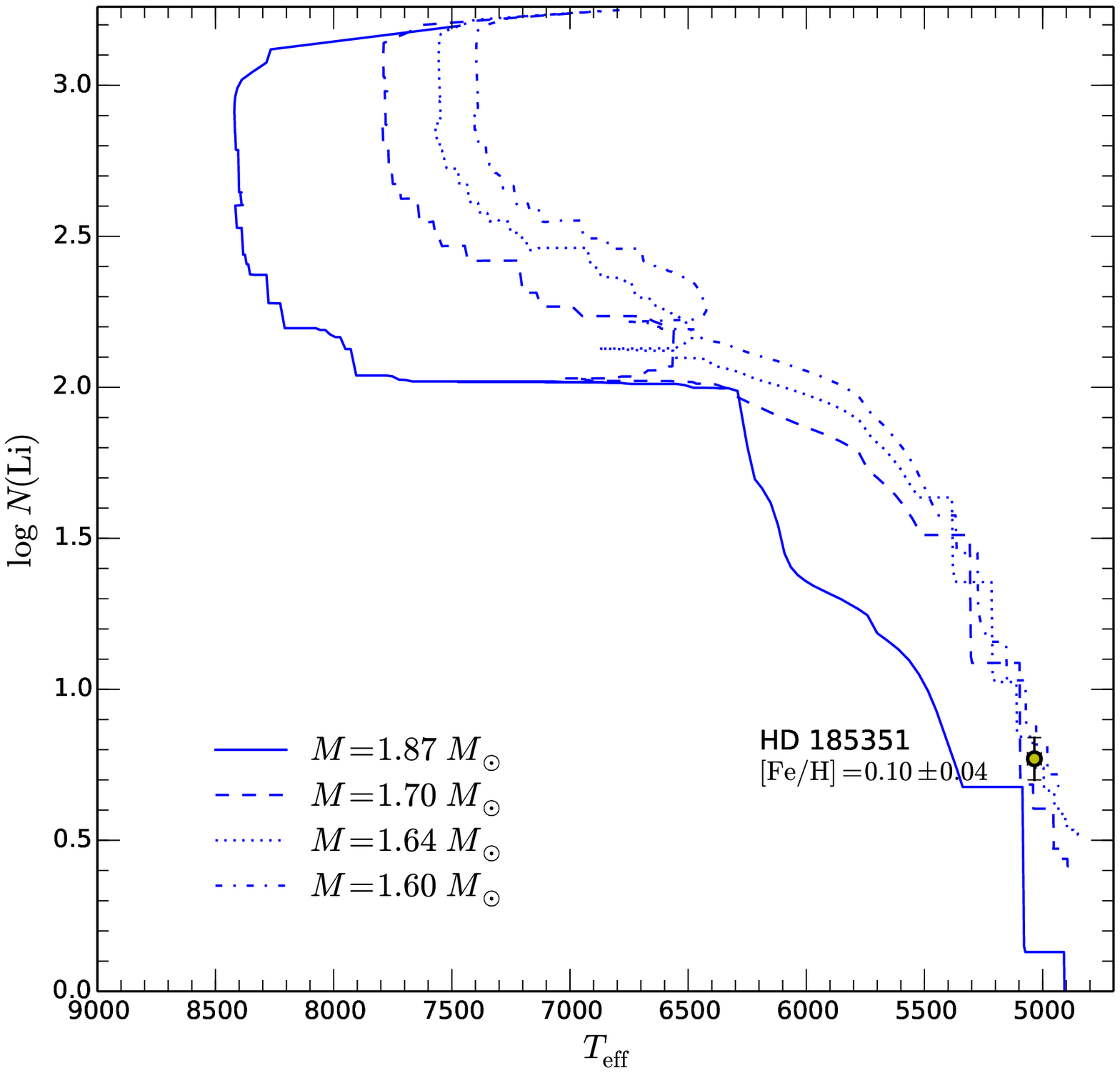}
\caption{Mass determination for HD 185351 based on its non-standard mixing history. 
The grid of models describing the evolution of the Li abundance (blue lines) was 
computed with the Toulouse-Geneva code \citep{2008Ap&SS.316...55H} for [Fe/H] = 0.10 
(from spectroscopy) and different masses. Interpolation of the observed values of 
$T_{\rm eff}$ and A(Li) (also from spectroscopy; yellow point) yielded the mass 
M = 1.64 $\pm$ 0.06 M$_{\odot}$.}
\label{models_lithium}
\end{figure}

We measured the physical parameters of HD 185351
using the atmospheric parameters $T_{\rm eff}$ and [Fe/H], the V
magnitude (5.18 $\pm$ 0.01) from the Hipparcos catalog \citep{1997yCat.1239....0E}, the
revised Hipparcos parallax (24.49 $\pm$ 0.22 mas; \citealt{2007ASSL..350.....V}), 
and the PARAM web interface\footnote{http://stev.oapd.inaf.it/cgi-bin/param} with the
PARSEC evolutionary tracks \citep{2012MNRAS.427..127B}.
The results were: M = 1.77 $\pm$ 0.04 M$_{\odot}$, R = 4.85 $\pm$ 0.11 R$_{\odot}$,
$\log$ g = 3.29 $\pm$ 0.02, and Age = 1.79 $\pm$ 0.09 Gyr.
Following the analysis of \citet{2009A&A...501..687D}, we also obtained an independent
estimate of the stellar mass from the analysis of the non-standard mixing history 
of HD 185351: M = 1.64 $\pm$ 0.06 M$_{\odot}$ (see Figure \ref{models_lithium}).
The independently measured masses are not only consistent, but also lie within the interval
($\approx$1.6 $-$ 2.0 M$_{\odot}$) defined by the extensive analysis of \citet{2014arXiv1407.2329J},
which includes results from interferometry, spectroscopy, and asteroseismology (see their Figure 7).

%
%

\section{Conclusions}

The mass of HD 185351 was determined using different methods and all estimated
values are higher than 1.5 M$_{\odot}$. These measurements are consistent with HD 185351
being the evolved counterpart of an A dwarf, which supports the results from
\citet{2010PASP..122..905J}. We plan to extend the spectroscopic analysis described here to
the entire sample of $\approx$300 evolved stars.

\acknowledgments{
LG would like to thank the financial support from CAPES, Ci\^encia
sem Fronteiras, Harvard College Observatory, and Funda\c{c}\~ao Lemann.
}

\end{document}